\documentclass[aps,apl,twocolumn,superscriptaddress,groupedaddress]{revtex4}
\usepackage{graphicx}
\usepackage{color}

\begin{document}
\title{Thermally-Assisted Spin-Transfer Torque Magnetization Reversal in Uniaxial Nanomagnets}
\author{D.~Pinna}
\email{daniele.pinna@nyu.edu}
\affiliation{Department of Physics, New York University, New York, NY 10003, USA}
\author{Aditi Mitra}
\affiliation{Department of Physics, New York University, New York, NY 10003, USA}
\author{D.~L.~Stein}
\affiliation{Department of Physics, New York University, New York, NY 10003, USA}
\affiliation{Courant Institute of Mathematical Sciences, New York University, New
             York, NY 10012, USA}
\author{A.~D.~Kent}
\affiliation{Department of Physics, New York University, New York, NY 10003, USA}

\begin{abstract}
We simulate the stochastic Landau-Lifshitz-Gilbert (LLG) dynamics of a uniaxial nanomagnet
out to sub-millisecond timescales using a graphical processing unit based micromagnetic code and determine the effect of geometrical tilts between the spin-current and uniaxial anisotropy axes on the thermally assisted reversal dynamics. The asymptotic behavior of the switching time ($I\rightarrow 0$, $\langle\tau\rangle\propto\exp(-\xi(1-I)^2)$) is approached gradually, indicating a broad crossover regime between ballistic and thermally assisted spin transfer reversal. Interestingly, the mean switching time is shown to be nearly independent of the angle between the spin current and magnet's uniaxial axes. These results have important implications for modeling the energetics of thermally assisted magnetization reversal of spin transfer magnetic random access memory bit cells.
\end{abstract}

\maketitle

A spin-polarized current passing through a small magnetic conductor is known to deposit its spin-angular momentum into the magnetic system~\cite{Slon,Berger}. This in turn causes the magnetic moment to precess and in some cases even switch direction. This has led to sweeping advances in the field of spintronics through the development and study of spin-valves and magnetic tunnel junctions (see, for example,~\cite{Brataas}). A thorough understanding of these phenomena, however, requires taking into account the effect of thermal fluctuations inducing diffusion in the moment of a magnetic material. This is of particular experimental relevance since spin-transfer effects on nanomagnets are often conducted at low currents, where noise is expected to dominate. Recent debate in the literature over the proper exponential scaling behavior between mean switching time and current elucidates how the thermally assisted properties of even the simplest magnetic setups leave much to be understood~\cite{Katine,Sun,Apalkov,LiZhang,Taniguchi}. By simulating reversal dynamics of a uniaxial macrospin model subject to spin-torque, we demonstrate how harnessing the vast computational parallelization capabilities intrinsic in Graphics Processing Unit (GPU) technology for numerical modeling can allow a deeper probing of the thermally activated regime. This leads us to a further understanding of the energetics at play in small magnetic systems subject to spin-torque effects.

The standard theoretical approach used to study the dynamics of small magnetic structures has been 
to treat the magnetic system as a macrospin in the spirit of Brown~\cite{Brown}, thus modeling the dynamics as those of a magnetic moment undergoing damped precession under the action of an effective field. After Slonczewski~\cite{Slon}, one assumes that the magnetic body absorbs spin-angular momentum perpendicularly to its orientation. As such, spin-torque effects are generally taken into account phenomenologically by modifying the macrospin's Landau-Lifshitz-Gilbert (LLG) dynamical equation with the addition of the spin-torque term. The LLG equation for a 
then reads:

\begin{eqnarray}
\mathbf{\dot{m}}&=&-\gamma'\mathbf{m}\times\mathbf{H}_\mathrm{eff}-\alpha\gamma'\mathbf{m}\times\left(\mathbf{m}\times\mathbf{H}_\mathrm{eff}\right)\nonumber\\
&-&\gamma' s\mathbf{m}\times\left(\mathbf{m}\times\mathbf{\hat{n}}_p\right)+\gamma'\alpha s\mathbf{m}\times\mathbf{\hat{n}}_p,
\end{eqnarray}
where $\gamma'=\gamma/(1+\alpha^2)$, $\gamma$ is the gyromagnetic ratio, $\alpha$ the Landau damping coefficient and $s=(\hbar/2e)\eta J$ is the spin-angular momentum deposited per unit time with $\eta = (J_{\uparrow}-J_{\downarrow})/(J_{\uparrow}+J_{\downarrow})$, the spin-polarization factor of incident current $J$. The last two terms describe a vector torque generated by current polarized in the direction $\mathbf{\hat{n}}_p$. The effective field $\mathbf{H}_\mathrm{eff}$ is immediately obtained once the energy landscape of the magnetic sample is specified. In the simplified case of a uniaxial monodomain magnet, it is characterized by the projection of the magnetization onto the uniaxial anistropy axis:

\begin{equation}
U(\mathbf{m})=-K(\mathbf{m}\cdot\mathbf{\hat{n}}_K)^2 - \mathbf{m}\cdot\mathbf{H},
\end{equation}
with $K=(1/2)M_SH_KV$ the uniaxial anisotropy of a magnet of volume $V$ and anisotropy field $H_K$, saturation magnetization $M_S$, applied field $\mathbf{H}$, and $\mathbf{\hat{n}}_K$ the unit vector representing the orientation of the uniaxial anisotropy (easy) axis.

Of special interest is the interplay between spin-torque and thermal noise effects. A spin current also induces magnetization fluctuations. But for typical experimental situations this noise is far smaller than the thermal noise, which we discuss below. Starting from (1), thermal fluctuations are modeled by considering the effect of a random applied magnetic field with mean zero and variance enforced by the fluctuation dissipation theorem. Doing this leads to a stochastic LLG equation with multiplicative noise~\cite{Brown,Palacios}. Experimentally, one is generally interested in the magnetization reversal properties of a thin film magnet. For currents less than a threshold current (to be defined below), the switching in such magnetic systems appears to be dominated by thermal-activation related reversal processes. Fitting experimental data
to theory is complicated by several factors. First, the validity of the macrospin model is questionable for samples typically studied (see, for example, \cite{Bedau}), which are so far larger than the exchange length of transition metals magnets ($\sim 5$ nm). Second, even within the macrospin approximation it is not clear what the relevant energetics are for thermally assisted spin transfer torque reversal. This is a consequence of the inherently non-conservative nature of the added spin-torque term which does not allow the construction of a generalized energy landscape suitable for Kramer's theory analysis. 

The energetics of thermally assisted reversal in the presence of spin transfer torques for a macrospin is the focus our analysis.
Here, theoretical progress has been hindered by the computational power needed to run numerical simulations to the desired degree of accuracy. The LLG equation, modified as a set of coupled stochastic equations can be studied in one of two ways: either by concentrating on its associated Fokker-Planck (FP) equation or by constructing a stochastic Langevin integrator to gather sufficient statistics on reversal. In either case simulations have thus far been unable to extrapolate to long enough times to approach the thermally activated regime. If on one hand, it is hard to maintain convergence of direct numerical solutions of the FP equation out to long timescales, gathering sufficient statistics to make proper use of a Langevin integrator has proven to be just as difficult a task. A recent paper by Taniguchi and Imamura ~\cite{Taniguchi}, suggests that the proper scaling behavior between mean switching time and current should be $\log(\langle\tau\rangle)\propto-(1-I)^2$, with $I=s/(\alpha H_K)$ the normalized current. This contradicts previous work proposing a proportionality to $-(1-I)$~\cite{Katine,LiZhang,Apalkov}. Nonetheless, numerical simulations have not yet been able to settle this disagreement.

To elucidate the most basic properties of thermally-assisted magnetization reversal under spin-transfer torque, we choose to consider a magnetic system solely characterized by its uniaxial anisotropy and focus on its thermally assisted dynamics in the absence of any applied fields. The energy landscape is given by the first term on the right hand side of (2). The equilibrium states in this landscape are simply given by parallel and anti-parallel configurations of the magnetization along the uniaxial anisotropy axis.  We consider our magnetic sample to be initially magnetized in thermal equilibrium with one of its two possible stable states (i.e. $m=+\mathbf{\hat{n}}_K$ or $m=-\mathbf{\hat{n}}_K$). Upon turning on a (positive) current, spin-torque effects will attempt to drive the magnetization toward aligning with the polarization axis $\mathbf{\hat{n}}_P$. If the current is then switched off again, the magnetization will relax toward the easy axis once more. To capture this behavior, it is very helpful to write down the dynamical equation for the projection $q\equiv\mathbf{m}\cdot\mathbf{\hat{n}}_K$ of the magnetization along the easy axis. For temperatures such that $\xi\equiv K/kT\gg1$, upon turning off the current, the sign of $q$ will specify to a high degree of accuracy in which state the magnetization will relax. We call this the ``projectional dynamics'', which read:

\begin{figure}[t]
\centering
\includegraphics[width=1.0\columnwidth]{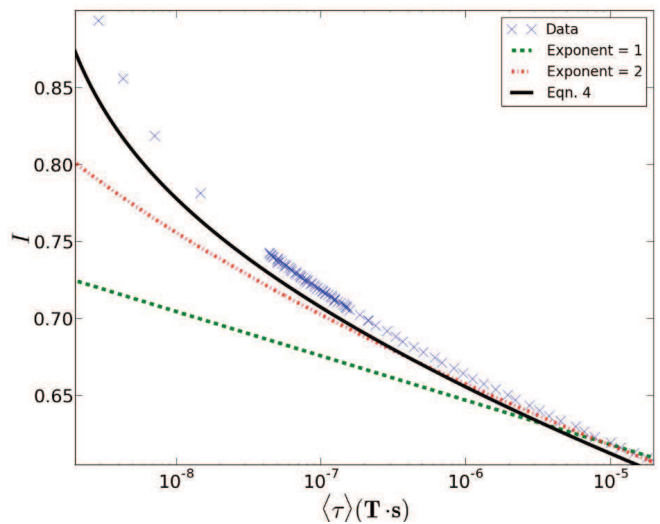}
 \caption{Mean switching time versus current in the sub-critical low current regime ($I<1$). Times are shown in units of ($s\cdot T$) where $T$ stands for Tesla: real time is obtained upon division by $H_K$. The red and green line are obtained by fitting the long time data to the functional form $\langle\tau\rangle=C\exp(-\xi(1-I)^{\mu})$, where $C$ is fitted numerically and $\mu$ is an exponent, either $1$ or $2$. The red curve ($\mu=2$) can be seen to approximate the numerical data asymptotically better than the green curve ($\mu=1$). The theoretical curve (in light blue) represents the asymptotic switching scaling given by Eqn. 4. The discrepancy between numerical data and theoretical asymptotic scaling of the mean switching time seems to indicate a broad crossover regime between ballistic and thermally assisted dynamics.}
 \label{FIG1}
\end{figure}

\begin{eqnarray}
\dot{q}&=&\alpha\left[(n_zI+q)(1-q^2)+n_zIq(q-\frac{m_z}{n_z})\right]\nonumber\\
&+&\alpha^2In_xm_y+\sqrt{\frac{\alpha}{\xi}(1-q^2)}\circ\dot{W}.
\end{eqnarray}
For definitiness, we have conventionally chosen the z-axis to be collinear with the polarizer (${\hat{n}}_p=\hat{z}$) and the easy axis to be lying in the z-x plane. To simplify the notation, then, $n_x$ and $n_z$ are simply the components of $\mathbf{\hat{n}}_K$. Furthermore, a natural time unit $\tau=\gamma'H_K t$ has been introduced. The last term appearing in the expression models the effects of thermal noise whose stochastic contribution is intended according to Stratonovich calculus (indicated by the symbol $\circ$) and $\dot{W}$ is a standard mean $0$, variance $1$ Wiener process~\cite{Karatsas}.  

In the simplified collinear case, where polarizer and uniaxial axes are aligned, the projectional dynamics are the same as the regular dynamics for the $m_z$ component of the magnetization, and are decoupled from both $m_x$ and $m_y$. The problem is effectively reduced to that of a 1D stochastic differential equation characterizing the switching behavior over all possible currents. It is easily seen that $I=1$ behaves as a crossover value above which all dynamics prefer the parallel state independently of the noise, and this value of $I$ is hereafter denoted as the ``critical current''. Analogously, the thermal regime is characterized by sub-critical current values ($I\ll 1$). The
thermally assisted properties are then easily understood by studying the mean first passage time over the effective barrier. This is done by contructing and solving the associated adjoint equation to the Fokker-Planck operator~\cite{Zwanzig, Kubo}. In the limit of low noise  ($\xi\gg 1$) the scaling behavior between mean switching time and current for sub-critical currents becomes:

\begin{equation}
\langle\tau\rangle\simeq\frac{\sqrt{\pi}}{\alpha}\frac{F[\sqrt{\xi}(1-I)] \exp[\xi(1-I)^2]}{1-I^2},
\end{equation}     
where $F[x]=\exp(-x^2)\int_0^x\exp(y^2)dy$ is Dawson's integral.  

To verify this result, we numerically solved the full stochastic LLG equation (Eqn. 1) by using a standard second order Heun scheme to ensure proper convergence to the Stratonovich calculus~\cite{Rumelin}. As mentioned above, the nanomagnet is initially in thermal equilibrium at a temperature $T$ in one of its two possible stable states at the start of the simulation ($\tau=0$), when the current is applied. At each time step, the random kicks are generated with strength given by

\begin{equation}
D=\frac{\alpha k_BT}{2K(1+\alpha^2)}=\frac{\alpha}{2(1+\alpha^2)\xi}
\end{equation}
which is obtained by imposing the fluctuation-dissipation theorem in the absence of applied currents. Such a diffusion constant is valid as long as the magnetic sample is in equilibrium with the thermal bath and effects due to Joule heating are neglected.

Statistics were gathered from an ensemble of $5000$ events with a step size of $0.01$ in reduced unit of time. The Landau damping constant is set at $\alpha=0.04$. Although the main results in this paper are shown for a barrier height of $\xi=80$, different barrier heights were explored. To solve the numerical equations out to long time regimes, the event dynamics were simulated in parallel on an NVidia Tesla C2050 graphics card. The large number of necessary random numbers were generated by employing a combination of the three-component combined Tausworthe ``taus88''~\cite{Ecuyer} and the 32-bit ``Quick and Dirty'' LCG~\cite{Press} whose statistical properties are discussed in the literature~\cite{Nguyen}. This hybrid generator provides an overall period of $2^{121}$. Comparison of our result (4) with numerical data is shown in Fig.~\ref{FIG1}: a $\langle\tau\rangle\propto\exp(-\xi(1-I)^2)$ is a much better fit to the data at long time scales than $\langle\tau\rangle\propto\exp(-\xi(1-I))$. Nonetheless, even at mean switching time scales four orders of magnitude larger than switching times at critical current, full asymptoticity in the numerics is still not reached exactly. This would seem to imply that the crossover regime, a regime characterized by  an interplay between deterministic and thermal effects, spans over very large mean switching timescales.

\begin{figure}
\centering
\includegraphics[width=1.0\columnwidth]{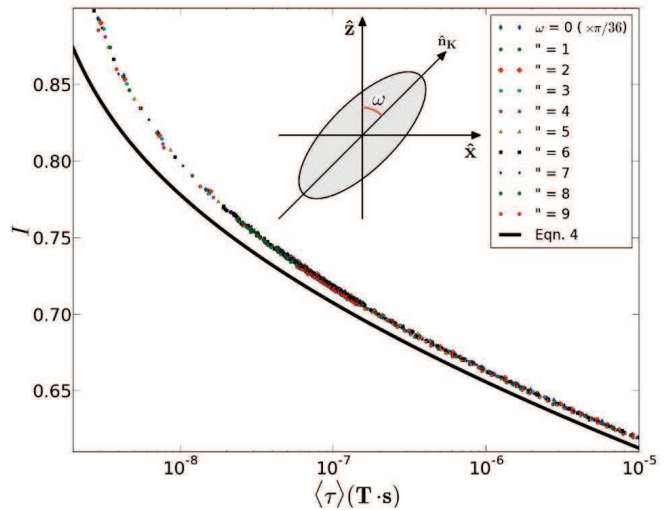}
  \caption{Comparison of the mean switching time versus current in the sub-critical low current regime for different values of angular separation between polarizer and easy axes, $\omega$, ranging from $0$ to $\pi/4$. All critical currents have been rescaled to $1$. Times are shown in units of ($s\cdot T$) where $T$ stands for Tesla: real time is obtained upon division by $H_K$. It is apparent that, up to a current rescaling, the mean switching time vs applied current is independent of $\omega$.}
  \label{FIG2}
\end{figure}

We now consider how breaking the collinear symmetry impacts the scaling behavior. In particular, (3) no longer provides a simplification, because it depends explicitly on both $m_z$ and $m_y$. These would need to be solved separately, given that $q$ is now a third variable. Nonetheless, in the thermally dominated
regime one can still make some useful approximations. For small values of $\alpha$, the term in square brackets is of leading order over the second ballistic term depending on $m_y$. This allows us, in the small $\alpha$ regime, to neglect the second ballistic term altogether. Furthermore, for low
sub-critical currents, we expect an initially anti-parallel configuration to diffuse not very far from its local energy minima and establish a transient equilibrium until a sufficiently strong thermal kick drives it over the effective barrier over timescales much smaller than those due to the deterministic spin-torque dynamics. This notion can be used to eliminate the $m_z$ dependence appearing in (3) by imposing its average value over constant energy trajectories. These in turn are nearly circular orbits around the uniaxial anisotropy axis for which $\langle m_z\rangle=n_zq$. Doing this leads the second term in the square brackets to vanish until the onset of the switching event. We can then think of (3), in such a limit, to behave in the following approximate form:

\begin{equation}
\dot{q}=\alpha(n_zI+q)(1-q^2)+\sqrt{\frac{\alpha}{\xi}(1-q^2)}\circ\dot{W}.
\end{equation}

This is reminiscent of the 1D projectional dynamics discussed in relation to the collinear limit. The only difference is the substitution $I \rightarrow n_zI$. Furthermore, analogously to the collinear limit, one can again define a critical current $I_c$ as the current above which all initial magnetic configurations will switch deterministically. This has already been derived elsewhere in the literature~\cite{Sun} and can be shown to give $I_c=1/cos(\omega)$, where $\omega$ is the angular tilt between uniaxial and polarizer axes. This in turn automatically implies that $n_zI=I/I_c$. In other words, the thermally activated dynamics are the same for all angular separations up to a rescaling by their respective critical current. We then expect that the mean switching time dependences remain functionally identical to the collinear case for all uniaxial tilts. This has been confirmed by comparison with data from our simulations, and the results are shown in Fig.~\ref{FIG2}. It is of interest to point out how the effect of angular tilt in thermally activated spin-transfer reversal is quite different from what one gets in studying the similar phenomenon of field driven reversal. It has been shown, in fact, that upon introducing an angular separation between easy and applied field axes, the scaling exponent changes abruptly in value from $2$ for the collinear case to $3/2$~\cite{Coffey}.  

These results have important implications for the analysis of experimental data in which measurements of the switching time versus current pulse amplitude are used to determine the energy barrier to magnetization reversal. Clearly use of the correct asymptotic scaling form is essential to properly determine the energy barrier to reversal. The energy barrier, in turn, is very important in assessing the thermal stability of magnetic states of thin film elements that are being developed for long term data storage in STT-MRAM. Further work should address how these results extend to systems with easy plane anisotropy and situations in which the nanomagnet has internal degrees of freedom, leading to a break down of the macrospin approximation.

We also note that current flow is a source of shot noise, which at low frequencies
acts like a white-noise source in much the same way as thermal noise. 
It is therefore interesting to understand when this additional source of noise plays a role.
For a magnetic layer coupled to unpolarized leads, the current induced noise on the magnetization dynamics was found to be 
$\frac{\Gamma_L/\Gamma_R}{(1+\Gamma_L/\Gamma_R)^2}V$~\cite{Mitra06}, where $V$ is the voltage drop across the magnetic layer, while $\Gamma_L/\Gamma_R$ is a dimensionless ratio characterizing the coupling strength of the magnetic layer to the left (L)
and right (R) leads. Thus the noise is maximal ($V/4$) for perfectly symmetrical couplings, and is smaller
in the limit of highly asymmetric contacts. This basic behavior, and the
order of magnitude of the effect, is not likely to be modified by polarized leads. 
We argue that the temperatures at which the experiments have been performed
current noise effects are not important.  The experiments are performed at room temperature where $T=300$ K. For an all metallic
device, such as a spin-valve nanopillar, the couplings are nearly symmetrical and at the critical current a typical voltage drop across the magnetic layer is less than $10\; \mu$V or, equivalently, $1$ K. For a magnetic tunnel junction device $V$ can be $\sim 1$ V. However, in this case the coupling is asymmetric. One lead (L) forms a magnetic tunnel junction with the nanomagnet, while the other (R) a metallic contact. This gives $\Gamma_R/\Gamma_L > 10^4$ and a relevant energy $\sim 1$ K, again far lower than room temperature.
It appears that current induced noise can only be important at room temperature for a nanomagnet coupled symmetrically between two tunnel barriers.

\begin{acknowledgments}
The authors would like to acknowledge A. MacFadyen and J. Z. Sun for many useful discussions and comments leading to this paper. This research was supported by NSF-DMR-100657 and PHY0965015.

\end{acknowledgments}

\end{document}